\documentclass[a4paper]{article}
\usepackage{caption}
\usepackage{multirow}
\usepackage{tikz}
\usepackage[affil-it]{authblk}
\usepackage{amssymb}
\usepackage{cite}

\title{Towards experiments with highly charged ions at HESR}

\author[1]{Rodolfo S\'anchez}
\author[1]{Angela Braeuning-Demian}
\author[1]{Jan Glorius}
\author[1]{Anton Kalinin}
\author[1]{Siegbert Hagmann}
\author[2]{Pierre-Michel Hillenbrand}
\author[1]{Yuri A. Litvinov}
\author[3]{Thomas K\"ohler}
\author[1]{Nikolaos Petridis}
\author[1]{Shahab Sanjari}
\author[1]{Uwe Spillmann}
\author[1,3,4]{Thomas St\"ohlker}

\affil[1]{GSI Helmholtzzentrum f\"ur Schwerionenforschung GmbH, Darmstadt, Germany.}
\affil[2]{Columbia University, New York City, USA}
\affil[3]{Friedrich-Schiller-Universität Jena, Jena, Germany}
\affil[4]{Helmholtz Institut Jena, Jena, Germany}

\date{}

\begin{document}

\maketitle

\begin{abstract}
The atomic physic collaboration SPARC is a part of the APPA pillar at the future Facility for Antiproton and Ion Research. It aims for atomic-physics research across virtually the full range of atomic matter. A research area of the atomic physics experiments is the study of the collision dynamics in strong electro-magnetic fields as well as the fundamental interactions between electrons and heavy nuclei at the HESR. Here we give a short overview about the central instruments for SPARC experiments at this storage ring.
\end{abstract}

\footnotetext{\textbf{Abbreviations:} HESR, High Energy Storage Ring; SPARC, Atomic Physics Collaboration; APPA, Atomic, Plasma Physics and Applications; FAIR, Facility for Antiproton and Ion Research}

\section{Introduction}\label{intro}

The construction of FAIR has speeded up considerably since 2017. This particle accelerator center will offer one of the highest intensities for relativistic beams of stable and unstable heavy nuclei. The SPARC collaboration aims for atomic physics experiments with highly-charged heavy ions (HCI) at the storage ring HESR at FAIR. The HESR, with a maximum magnetic rigidity of 50~Tm, will store cooled beams, e.g. hydrogen-like, helium-like and lithium-like ions, with energies up to 5~GeV/u. The comparably simple electronic structure of these ions provides ideal conditions for investigating the collision dynamics and atomic structure in strong electromagnetic fields, which are enhanced and/or become into reach due to the relativistic boost. Under these conditions radiative effects, like correlations, relativistic and QED, can be disclosed. 

Several precision experiments, which can be addressed at the HESR, have been proposed in the last years~\cite{Stoehlker2013, Stoehlker2015, Hagmann2013, Bosch2013, Litvinov2013, Gumberidze2015, Hillenbrand2015, Rothhardt2015, Noertershaeuser2015}. This include the studies of pair-production phenomena, negative continuum dielectronic recombination, relativistic photon-matter interaction, target ionization, bound state QED, nuclear structure via isotope shifts, exotic nuclear decay modes, parity non-conservation effects and field ionization of HCI. 

A feasibility study was performed in advance to investigate the operation of the HESR with highly charged ions~\cite{Stoehlker2012}. This was further investigated and study in detail in a more recent work~\cite{Kovalenko2015}.

To be able to perform the experiments mentioned above, additional instrumentation must be brought into the HESR. A brief description of this setup is given in what follows.

\section{High-energy storage ring}\label{hesr}

The optical lattice of this storage ring has been already described elsewhere~\cite{Stoehlker2013}. Its magnetic rigidity of 50~Tm will allow storing heavy highly charged ions up to U$^{92+}$ with energies varying in the range of 200-5000~MeV/u. It has two straight sections with a length of 132~m each and two 180$^{\circ}$ arcs with a length of 155.5~m each, making a total circumference of 575~m. Within each arc there are two straight sections, so called, “missing dipole” sections. Each of these has a length of about 4.4~m. For the antiproton operation these sections consist of a straight vacuum pipe terminated with “quick release” (QCF-100) flanges.  For the SPARC experiments the missing-dipole sections in the north-western and south-western arcs will be used, see section \ref{sparcs}. Their locations are shown in figure~\ref{fig:sshesr}. 	The optical lattice for heavy ion beam operation has been studied in detail in \cite{Kovalenko2015}. As a result, ion-beam emittances of $\epsilon_{x,y} = (0.25,0.15)$~mm~mrad for 10$^8$ stored uranium ions were obtained from the simulations. This corresponds to a radius of 4.2~mm for the 2-$\sigma$ beam, which satisfies very well the requirement for many of the planned atomic physics experiments.

\section{SPARC-setup at the missing dipole}\label{sparcs}
\begin{figure}[t]
\centerline{\includegraphics[width=\linewidth]{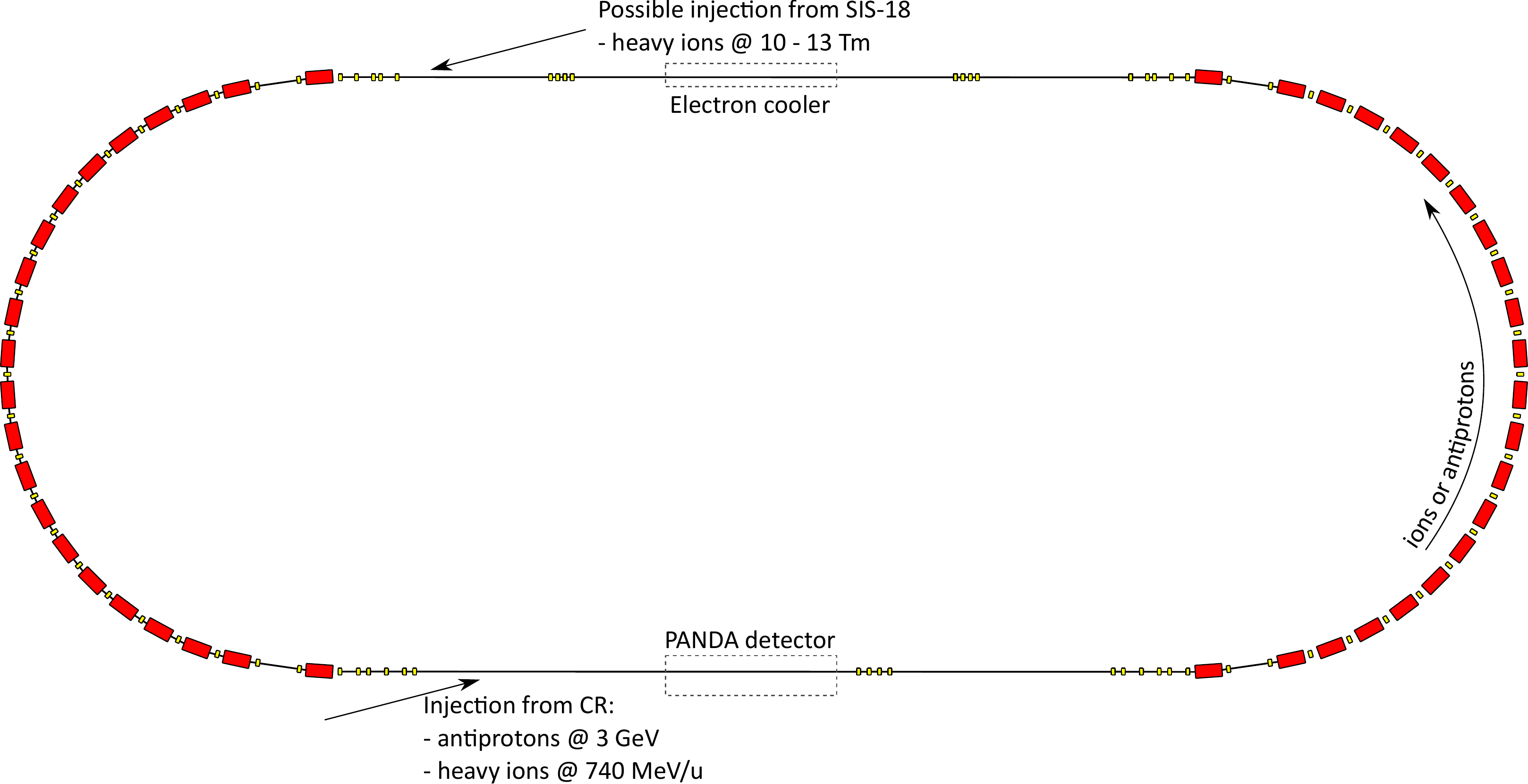}}
\caption{The HESR optical lattice showing only the dipoles (in red) and quadrupoles (in yellow). Either ions or antiprotons will be stored in counter-clockwise direction. For ion-beam operation the polarity of the magnets will be inverted. Inset: The SPARC-setup at the missing-dipole section(s). The main components are shown in different colors. The frame construction, which holds the entire setup is not shown. \label{fig:sshesr}}
\end{figure}

The inset in figure~\ref{fig:sshesr} shows a very simplified schematic view of the setup. This is the basic setup, which can be modified to meet the requirements of the different experiments. This section is quite similar to the one presently installed at the ESR and it has the goal to accommodate several experimental setups to run experiments in parallel. The main components are described in the following sections.

\subsection{Gas target}\label{gastrg}
This is the center piece of the setup. Many of the planned experiments will employ the interaction of fast moving heavy ions with the supersonic gas target. The realization of this device is based on experimental studies carried out at the ESR in order to meet the new experimental conditions at the HESR~\cite{Petridis2015}. The target beam is formed by expanding the desired gas (H2, He, N2, Ar, Kr, Xe) through a nozzle at defined source conditions into the vacuum. Target densities ranging from 10$^{10}$ up to 10$^{14}$~p/cm$^2$ are available and can be improved by at least one order of magnitude in new concept with:
\begin{itemize}
	\item	much lower source temperatures.
	\item	reduction of the distance between the nozzle and the interaction point.
	\item	the interaction length in the range of 1-5 mm is adjustable by employing exchangeable skimmers with different aperture diameters.
\end{itemize}

\subsection{Interaction chamber}\label{intcham}
Depending of which experiment will be carried out, different types of interactions chambers can, in principle, be built. Because this could also have consequences for the high-vacuum conditions, such a change will however rarely be made. This chamber has different flanges under different angles of observation. This offers many experimental possibilities. For example, it is possible to accommodate different experimental setups or detectors in order to run the experiments in parallel. A couple of ports for CCD cameras are also available in order to check the overlap between the ion-beam and gas-target. The technical design of this chamber for the HESR is currently under construction.

\subsection{Lepton spectrometer}\label{lepton}
The lepton-spectrometer setup is downstream after the gas target. This position was chosen for background-reduced detection of the reaction products. The technical parameters of this spectrometer have been determined by the signature of the respective collision processes under investigation, which are intended for the experimental study at the HESR. The immediate goal is the detection of the positrons emanating in forward direction from the interaction volume in collisions of HCI with atoms from the supersonic gas target. Examples of this kind of collision processes are the negative-continuum dielectronic recombination (NCDR), a process which has not been verified by experiment so far, and the study of the ion-induced pair production. A brief overview of the physics case is discussed in \cite{Hillenbrand2015, Hagmann2018}.
       
\subsection{Laser beamline}\label{laser}
For the coupling of laser beams, the vacuum beamline have to be extended outside the missing dipole region. For this purpose the dipole magnets were modified. The dipole coils were moved apart by a distance of 20 mm. The vacuum chambers inside the dipole magnets close to the SPARC-setup were also modified and an additional vacuum pipe CF16 was added, its axis is congruent with the missing-dipole beamline axis. The vacuum pipe for lasers ends with a flange just after leaving the dipole magnets. On the flange either an optical window or the differential pumping unit for the XUV laser \cite{Rothhardt2018} can be attached.  In this way the laser beams can be brought into the HESR and merged them with the ion-beam into the missing dipole sections. For the collinear- or the anticollinear excitation, the laser beams will be coupled in the South-Western arc or North-Western arc, respectively.  Thus the wavelengths of the laser and/or laser-driven source in the visible and XUV regime can be boosted in combined experiments with heavy ions. For example, E1 dipole transitions in the x-ray regime can be probed using a new developed XUV-laser source or off-the-shelf UV-lasers \cite{Rothhardt2015, Noertershaeuser2015}. For fluorescence detection, novel detectors, which have been already tested at the ESR, will be used \cite{Winzen2018}. The optical detection region at the SPARC setup is located downstream after the lepton spectrometer. 

\subsection{Detectors}\label{detector}
Various types of general purpose detectors and depending on the need of the future experiments, special detectors, are foreseen for installation at the HESR and have already been applied in experiments by the atomic physics group \cite{Anjelkovic2015}. For example
\begin{itemize}
	\item	Resonant Schottky pick-ups have proven to be indefensible tools in ion storage rings both for diagnostics and as experimental detectors~\cite{Nolden2011}.
	\item Si(Li)-and Ge(i)-Compton polarimeters for x-ray spectroscopy and x-ray polarimetry~\cite{Spillmann2018, Vockert2018}.
	\item Silicon microcalorimeters for high-precision X-ray spectroscopy~\cite{Kraft-Bermuth2018}.
	\item Roman pots/pockets with the particle detectors for the detection of the reaction products after the interaction of the HCI with the gas-target~\cite{Kovalenko2015}. These detectors will be installed in the so-called pump chambers at the HESR. A technical design of the pocket, which suits the pump chamber is presently under construction. 
\end{itemize}


\end{document}